\font\eightrm=cmr8 at 8pt
\documentclass{oxarticle}
\usepackage{amsmath, amssymb}
\usepackage{graphics, setspace}
\usepackage{epstopdf}
\usepackage{bm}
\usepackage{tikz}
\usepackage{xcolor}

\newcommand{\oCt}{{\rm (\oC \x^2 \oC)  }}

\newcommand{\XM}{{Exotic Model}}

\newcommand{\ME}{Master Equation}

\newcommand{\cM}{{\cal M}}

\newcommand{\CDSS}{{Chiral Dotted Spinor Superfield}}

\newcommand{\cE}{{\cal E}}

\newcommand{\cA}{{\cal A}}

\newcommand{\Exists}{\bm{\exists}\kern-0.6em\bm{\exists}}

\newcommand{\EI}{Exotic Invariant}
\newcommand{\ei}{exotic invariant}

\newcommand{\SSM}{Supersymmetric Standard Model}

\newcommand{\bt}{\begin{tabular}{c}}
\newcommand{\et}{\end{tabular}}

\newcommand{\eb}{\ee\be } 
\newcommand{\ebp}{\rt.\ee\be\lt.} 
\newcommand{\ebpp}{\rt.\rt.\ee\be\lt.\lt.} 
 
\newcommand{\bmat}{\lt ( \begin{array} }
\newcommand{\emat}{  \end{array} \rt )}

\newcommand{\oP}{{\ov P}}

\newcommand{\oE}{{\ov E}}

\newcommand{\oa}{{\ov a}}

\newcommand{\oY}{{\ov Y}}
\newcommand{\og}{{\ov g}}
\newcommand{\oy}{{\ov \y}}

\newcommand{\om}{{\ov m}}

\newcommand{\oC}{{\ov C}}

\newcommand{\oF}{{\ov F}}
\newcommand{\A}{{\ov A}}

\renewcommand{\a}{\alpha}	
\renewcommand{\b}{\beta}
\newcommand{\g}{\gamma}
\renewcommand{\d}{\delta}

\newcommand{\m}{\mu}
	
\newcommand{\x}{\xi}

\newcommand{\f}{\phi}
\renewcommand{\c}{\chi}
\newcommand{\y}{\psi}

\newcommand{\G}{\Gamma}

\newcommand{\Lam}{\Lambda}

\renewcommand{\S}{\Sigma}

\newcommand{\la}{\label}
\newcommand{\ci}{\cite}

\newcommand{\ds}{\documentstyle}	
\newcommand{\fr}{\frac}

\newcommand{\pa}{\partial}
\newcommand{\ov}{\overline}
\newcommand{\br}{\begin{rant}}
\newcommand{\er}{\end{rant}}
\newcommand{\beC}{\begin{Conjecture}}
\newcommand{\eeC}{\end{Conjecture}}
\newcommand{\be}{\begin{equation}}
\newcommand{\ee}{\end{equation}}
\newcommand{\ba}{\begin{array}} 
\newcommand{\ea}{\end{array}}
\newcommand{\bea}{\begin{eqnarray}}
\newcommand{\eea}{\end{eqnarray}}
\newcommand{\ra}{\rightarrow}

\newcommand{\Lra}{\Leftrightarrow}
\newcommand{\lt}{\left}
\newcommand{\rt}{\right}

\newcommand{\ben}{\begin{enumerate}}
\newcommand{\een}{\end{enumerate}}
\newcommand{\bitem}{\begin{itemize}}
\newcommand{\eitem}{\end{itemize}}

\newcounter{orange} 
\newcounter{apple} 
\addtocounter{apple}{1}
\newcounter{grape} 
\addtocounter{grape}{1}
\newcommand{\numberhere}{Feb4\;}
\newcommand{\articlenumber}{E4FINAL.tex}

\proofmodefalse
\usepackage{color}

\newcommand{\mathsym}[1]{{}}
\newcommand{\unicode}[1]{{}}

\usepackage{tikz}

\begin{document}

 \begin{center}

{ \Large   The EP Model and its Completion Terms (E4)  \\[1cm]}  
%

\renewcommand{\thefootnote}{\fnsymbol{footnote}}

{\Large John A. Dixon\footnote{jadixg@gmail.com, john.dixon@ucalgary.ca}\\Physics Dept\\University of Calgary 
\\[1cm]}  
\end{center} 
\Large
 
 \begin{center}Abstract
\end{center}
\Large
 Here we present the simple example of an \EI\ with just two  chiral electron supermultiplets ${\widehat E}$ and ${\widehat P}$. In this  example we include a mass term, and that means that there is a constraint on the \EI.  The constraint is easily solved for this simple case. Here we also exhibit a simple conjecture for the Completion Terms.    This simple example is very useful, because the constraint that arises in the case of the XM, in E6,  is just as easy to solve, and the Completion Terms there are also very similar to those here.  So this simple EP model is very useful for understanding the XM, which is what results from adding an \EI\ to the rather complicated \SSM. 

 \section{Introduction}
 \Large

In this paper we write down explicitly a simple model with the two parts of an electron plus its superpartners, an electron mass term, and an exotic invariant made from these fields.  This paper E4 is the fourth paper in the E series \ci{E1}--\ci{E??}. This paper and its successor E5 are simple models to prepare for the  XM model in E6.

\section{The  Action $\cA$ for this theory}
\la{actionsection}

The action here is only a little more complicated than the one in \ci{E3}.  Here is the action:
\be
\cA= \cA_{\rm Fields}+\cA_{\rm PseudoFields}
\la{sumfofieldandpseudo}
\ee

\subsection{Field parts $\cA_{\rm Fields}$ of the action}
The part $\cA_{\rm Fields}$ in (\ref{sumfofieldandpseudo}) has the form

\be
\cA_{\rm Fields}= {\cal A}_{\rm E} + {\cal A}_{\rm P} +\cA_{\rm CDSS,\;Kinetic}   
  +{\cal \A}_{\rm CDSS,\;Chiral} +\cA_{\rm CDSS,\;Chiral} \eb +{\cal A}_{\rm Mass}+{\cal \A}_{\rm Mass}  
  \ee
where the following two terms are the same as in E3, except there are now two of them:
\be {\cal A}_{\rm E} = 
a_0 \int d^{4}x \left \{
  F_E    {\ov F}_{E}
  -
\y_E^{ \a  }  \pa_{ \a \dot \a  }  {\ov \y}_{ E}^{  \dot \a}  
-  \pa_{ \m  }
  E   \pa^{ \m  }  \oE  
\rt\}
\la{Ekineticaction}\ee

\be
 {\cal A}_{\rm P} = 
a_1 \int d^{4}x \left \{
  F_P    {\ov F}_{P}
   -
\y_P^{ \a  }  \pa_{ \a \dot \a  }  {\ov \y}_{ P}^{  \dot \a}  
-  \pa_{ \m  }
  P   \pa^{ \m  }  
    \oP  
\rt\}
\la{Pkineticaction}\ee
The following three terms have not changed at all from E3:
 \be
\cA_{\rm CDSS,\;Kinetic}    
= a_2 \int d^4 x   
\lt \{  {\f}_{\dot \b} \pa^{\a \dot \b} \Box {\ov \f}_{\a} 
+ {W}_{ \a \dot \b } \pa^{\a \dot \d} \pa^{\g \dot \b}  
 {\ov W}_{\g \dot \d} 
+  {\c}_{\dot \b} \pa^{\a \dot \b}  {\ov \c}_{\a}  
\rt \} 
\la{Phikineticaction}\ee

\be \cA_{\rm CDSS, \;Chiral}  =    a_3  \int d^4 x   
\lt \{2  \f^{\dot \b} \Box   \c_{\dot \b}  
 +    X^{ \dot \b \a}   \Box     X_{\a \dot \b }
\rt \} 
\ee

\be {\ov \cA}_{\rm CDSS, \;Chiral}  =  \oa_3  \int d^4 x   
\lt \{2  {\ov \f}^{ \b} \Box   {\ov \c}_{ \b}  
 +   {\ov  W}^{\b  \dot \a} \Box     {\ov  W}_{  \b  \dot \a }
\rt \} 
\ee
Now there is a complex chiral scalar term here in E4, whereas in  E3 we had  no chiral term for the chiral scalar supermultiplet.  This arises, as usual in SUSY theories, from a superpotential:

\be  
 {\cal A}_{\rm Mass} =m \int d^{4}x \left \{
   F_P   E+  F_E   P -  \y_P^{\a}    \y_{E \a} 
\rt\}
\la{EPmassaction} 
\ee

\be {\cal \A}_{\rm Mass} = \om
\int d^{4}x \left \{
  \oF_P   \oE+   \oF_E   \oP-    \oy_P^{\dot \a}    \oy_{E \dot \a} 
\rt\}
\la{CCEPmassaction} 
\ee
 
The terms (\ref{EPmassaction}) and (\ref{CCEPmassaction}) make an important difference for the \EI, because now we need to solve the constraint equation that always arises when there is a superpotential.

\subsection{PseudoField parts $\cA_{\rm PseudoFields}$ of the action}
The part $\cA_{\rm PseudoFields}$ in 
(\ref{sumfofieldandpseudo}) has a form appropriate for the terms in $\cA_{\rm Fields}$.  This is very similar to the terms in E3:

\be
\cA_{\rm PseudoFields}= {\cal A}_{\rm PseudoFields,E} + {\cal A}_{\rm  PseudoFields,P }
+\cA_{\rm  PseudoFields,CDSS} \eb
+{\rm Complex\;Conjugates}
 + {\cal A}_{\rm Structure} \ee
where
\be
{\cal A}_{\rm PseudoFields, E}  = \int d^4 x \;
\lt\{ \G_{E}    \lt ( \y_{ E \b} {  C}^{  \b} 
+ \x^{\nu} \partial_{\nu}  E \rt )
\ebp +
 Y_{E}^{ \a}
 \lt ( \pa_{ \a \dot \a  }  E \oC^{\dot\a}
 + F_{E} {C}_{ \a}
+ \x^{\nu} \partial_{\nu}  \y_{ E \a}
\rt )+
\Lam_{E}  \lt ( {\ov C}^{\dot \a}
 \pa_{ \a \dot \a  }  {  \y}_{E}^{  \a}
 + \x^{\nu} \partial_{\nu}   F_E   
\rt )
\rt \}
\ee
\be
{\cal A}_{\rm PseudoFields, P} = \int d^4 x \;
\lt \{
 \G_{P}    \lt ( \y_{ P \b} {  C}^{  \b} 
+ \x^{\nu} \partial_{\nu}  P\rt )
\ebp
+
 Y_{P}^{ \a}
 \lt ( \pa_{ \a \dot \a  }  P
 + F_{P} {C}_{ \a}
+ \x^{\nu} \partial_{\nu}  \y_{ P \a}
\rt )+
\Lam_{p}  \lt ( {\ov C}^{\dot \a}
 \pa_{ \a \dot \a  }  {  \y}_{P}^{  \a}
 + \x^{\nu} \partial_{\nu}   F_P   
\rt )
\rt \}\ee
\be
{\cal A}_{\rm PseudoFields, CDSS} = \int d^4 x \;
\lt \{ G^{\dot\a}\lt (  C^{\a} X_{ \a \dot \a}+
 \x^{\g \dot \d} \partial_{\g \dot \d}\f_{\dot \a}\rt )
 \ebp
+\S^{ \a  \dot \b}\lt (
\pa_{ \a \dot \g }  \f_{\dot\b} {\ov C}^{\dot \g}  
+ 
C_{\a}   
\c_{\dot\b}
+ \x^{\g \dot \d} \partial_{\g \dot \d}  X_{ \a \dot\b}
\rt )
+L^{\dot \a}\lt (
  \pa^{\b \dot \g}  X_{ \b  \dot \a}   {\ov C}_{\dot \g} 
+ \x^{\g \dot \d} \partial_{\g \dot \d}  \c_{\dot\a}
\rt ) \rt \}
 \ee
\be
 {\cal A}_{\rm Structure} =
 -K_{\a \dot \b} C^{\a} \oC^{\dot \b}
\ee

\section{The details of the \EI\ for this EP model}
\la{eisection}

This section is the analogue of section 3 in E3.  There need to be two, very similar, but different, pieces to this \EI\footnote{This expression comes from the spectral sequence term
 \be
\cE_{X}= \oCt \f^{\dot \a} \lt (  E \oy_{E \dot \a}-  P \oy_{P \dot \a}  \rt ) \in E_{\infty}
\ee
and the minus sign comes from the constraint which has the form
\be
d_2  \cE_{X}=\oCt
\f^{\dot \a} \lt (  m E P \oC_{\dot \a}-  m P E  \oC_{\dot \a} \rt )
=0
\ee  These spectral sequence issues are discussed elsewhere throughout this series, but here we ignore them for simplicity.  But the results here are all dependent on the results of E2.}:
\be
\cA_{\rm X} = \cA_{\rm X, E }  -  \cA_{\rm X, P } 
\la{theEI}\ee
where 
\be
 \cA_{\rm X, E } =  \int d^4 x  \lt \{  \f^{\dot \a}\lt [ \lt ( 
b_{1}E \G_{E} 
+ b_{2}\y_{E}^{\b} Y_{E \b}
+b_{3}\Lam_{E}  F_{E}
\rt ) \oC_{ \dot \a}
\ebpp + 
b_{4} \y_{E}^{\a}\pa_{\a \dot \a} \oE 
 + 
b_{5} F_{E}\oy_{E  \dot \a}\rt ]
+ X^{\b \dot \a } \lt [ \lt (b_{6}  E Y_{E \b} + b_{7} \y_{E \b} \Lam_{E}\rt )
\oC_{ \dot \a}+ 
b_{8} \y_{E \b}\oy_{E \dot \a} + 
\ebpp b_{9}{E} \pa_{\b \dot \a} \oE   \rt ]
  +  \c^{\dot \a} \lt [b_{10}  E \Lam_{E } \oC_{ \dot \a} +
b_{11}E\oy_{E\dot \a}
 \rt ] \rt \}  
\la{eiforE} 
\ee
and
\be
b_{1}
=
b_{2}=
b_{3}=
b_{4}=b_{7}=b_{10}=1; 
b_{5}=
b_{6}=
b_{8}=
b_{9}=
b_{11}= -1
\la{valsofb}
\ee
The  expression $\cA_{\rm X,  P} $ is identical to $\cA_{\rm X, E} $ except that  $E\ra P$.  The minus sign in (\ref{theEI}) is crucial, as we will see. In E3 we had only one term in (\ref{theEI}), because we had no terms that come from a superpotential. Here  in E4 we have the mass terms above in (\ref{EPmassaction}) and (\ref{CCEPmassaction}) and that forces us to take two terms in (\ref{theEI}), as we will now see.

\section{The coefficients $b_i, i =1\cdots 11$ in the \ei\  $ \cA_{\rm X} $}
\la{coeffsec}

These coefficients (\ref{valsofb}) are determined by taking a general form as shown and determining the coefficients $b_i, i =1\cdots 11$ by the equations
\be
\d \cA_{\rm X} =0
\la{constraintoncAX}
\ee
where $\d$ is the nilpotent BRS operator constructed from the \ME\ in this theory. It can be found in section (\ref{brsoperatorsection}).
   In fact we can calculate that \be
\d \cA_{\rm X,E} =  
\d \cA_{\rm X,P} =
\eb
\int d^4 x \lt \{
   -  \f^{\dot \a}   E   m  F_P    \oC_{ \dot \a} 
 - \f^{\dot \a}  F_{E}  \oC_{ \dot \a} m P  
-X^{ \a\dot \a}    E \oC_{ \dot \a}m
\y_{P \a} 
\la{variationsofXEandXP1}\ebp
 - X^{ \b\dot \a}   \y_{E \b} \oC_{ \dot \a} m P  
+   \f^{\dot \a} \y_{E}^{\a} \oC_{ \dot \a}m
\y_{P \a} 
-  \c^{\dot \a}    E \oC_{ \dot \a} m P  
\rt \}
\la{variationsofXEandXP2}\ee
This expression is symmetric when $E\Lra P$ and so, for the difference, we get
\be
\d \cA_{\rm X} =\d \lt \{\cA_{\rm X,E} -  
\cA_{\rm X,P}\rt \}=0
\ee
This is the reason for the constraint  in the spectral sequence, discussed and derived in \ci{E2}. 
There is always a constraint when there is a superpotential term in the action.  

The constraint requires us to find a set of signs so that the terms (\ref{variationsofXEandXP1}) and (\ref{variationsofXEandXP2}) cancel between the nearly identical parts of the \EI\ (\ref{theEI}). 
Exactly the same sort of thing happens in E6 for the XM, which couples the \EI\ for that case to the \SSM.  To verify the results of this section it is best to use a computer program, and we will return to that question in \ci{E10}.  Because this term arises from the considerations in E2, it is guaranteed that there is a non-zero solution for the equations here, because the spectral sequence states that there is indeed a non-zero term here.

\section{The \ME}
\la{mastereqsection}

The \ME\ is discussed in the references \ci{bagger}--\ci{dixmin}.  It has the following form here: 
\be
\cM_{\rm }=\cM_{\rm  E}+\cM_{\rm  P} 
+\cM_{\rm CDSS}+\cM_{\rm Structure}=0
\la{mastereq}
\ee
Again this is very similar to the discussion in \ci{E3}.
This is satisfied by the action $\cA$ 
in  (\ref{sumfofieldandpseudo}) in section \ref{actionsection}, as is well known from the usual considerations for these \ME s.  The action $\cA$  here that appears in the following terms is  that action $\cA$ 
in  (\ref{sumfofieldandpseudo}):
\[
\cM_{\rm  E}=
\int d^4 x \lt \{
\fr{\d \cA}{\d E} \fr{\d \cA}{\d \G_E}
+
\fr{\d \cA}{\d \y_{E \a} } \fr{\d \cA}{\d Y_{E}^{\a}}
+
\fr{\d \cA}{\d F_E} \fr{\d \cA}{\d \Lam_E}
+
\fr{\d \cA}{\d \oE} \fr{\d \cA}{\d \ov\G_E}
\rt.\]\be\lt.+
\fr{\d \cA}{\d \oy_{E \dot \a}} \fr{\d \cA}{\d \oY_E^{\dot \a}}
+
\fr{\d \cA}{\d \oF_E} \fr{\d \cA}{\d \ov\Lam_E}
\rt \}  
\la{mastereq1}
\ee
\[
\cM_{\rm  P}=
\int d^4 x \lt \{
\fr{\d \cA}{\d P} \fr{\d \cA}{\d \G_P}
+
\fr{\d \cA}{\d \y_{P\a}} \fr{\d \cA}{\d Y_{P}^{\a}}
+
\fr{\d \cA}{\d F_P} \fr{\d \cA}{\d \Lam_P}
+
\fr{\d \cA}{\d \oP} \fr{\d \cA}{\d \ov\G_P}
\rt.\] \be\lt.+
\fr{\d \cA}{\d \oy_{P \dot \a}} \fr{\d \cA}{\d \oY_P^{\dot \a}}
+
\fr{\d \cA}{\d \oF_P} \fr{\d \cA}{\d \ov\Lam_P}
\rt \}  
\la{mastereq2}\ee
\[
\cM_{\rm CDSS} =\int d^4 x \lt \{
\fr{\d \cA}{\d \f_{\dot\a}} \fr{\d \cA}{\d  G^{\dot\a}}
+
\fr{\d \cA}{\d X_{\a\dot\a}} \fr{\d \cA}{\d \S^{\a\dot\a}}
+
\fr{\d \cA}{\d\c_{\dot\a}} \fr{\d \cA}{\d L^{\dot\a}}
 +\fr{\d \cA}{\d \ov \f_{\a}} \fr{\d \cA}{\d  \ov G^{ \a}}
\rt.\]\be\lt.+
\fr{\d \cA}{\d \ov X_{\a\dot\a}} \fr{\d \cA}{\d \ov\S^{\a\dot\a}}
+
\fr{\d \cA}{\d\ov\c_{\a}} \fr{\d \cA}{\d \ov L^{\a}}
\la{mastereq3}
\rt \}  
\ee
\be
 \cM_{\rm Structure}=
\fr{\pa \cA}{\pa K_{\a \dot \b}} 
\fr{\pa \cA}{\pa \x^{\a \dot \b}}    
\la{masterstructure}
\ee

 \section{Tables of the Nilpotent BRS operator $\d$}
 \la{brsoperatorsection}
 Now we must take the `square root' of the various parts of the \ME\ to generate $\d$.  The action $\cA$  here is again $\cA$ 
in  (\ref{sumfofieldandpseudo}).
This results in the following:
\be
\la{brstransE} 
\vspace{.1in}
\framebox{{$\begin{array}{lll}  
& &{\rm Nilpotent  \;Transformations\; for\; E}\\
\d E&= & 
\fr{\d {\cal A}}{\d \G_E} 
=  \y_{E  \b} {C}^{  \b} 
+ \x^{\g \dot \d} \partial_{\g \dot \d} E
\\
\d {\ov E} &= & 
\fr{\d {\cal A}}{\d {\ov \G}_E} 
=  {\ov \y}_{E  \dot \b} {\ov C}^{ \dot  \b} 
+ \x^{\g \dot \d} \partial_{\g \dot \d} {\ov E}
\\

\d \y_{E \a} &  =& \fr{\d {\cal A}}{\d {  Y}_E^{   \a} } = 
\pa_{ \a \dot \b } E {\ov C}^{\dot \b}  
+ 
C_{\a}   
F_E
+ \x^{\g \dot \d} \partial_{\g \dot \d}  \y_{E\a  }
\\

\d
 {\ov \y}_{E \dot \a} &  =& 
\fr{\d {\cal A}}{\d { {\ov Y}}_E^{\dot   \a} } = 
\pa_{ \a \dot \a }  {\ov E}_{} {C}^{\a}  
+ 
{\ov C}_{\dot \a}   
{\ov F}_{E}
+ \x^{\g \dot \d} \partial_{\g \dot \d} 
 {\ov \y}_{E \dot \a} 
\\
 \d F_E 
&=&\fr{\d {\cal A}}{\d \Lam_E} = 
  \pa_{\a \dot \b}   \y_E^{ \a} {\ov C}^{\dot \b} 
+ \x^{\g \dot \d} \partial_{\g \dot \d}  F_E 
\\
 \d \oF_E 
&=&\fr{\d {\cal A}}{\d \ov\Lam_E} = 
  \pa_{\b \dot \a}   \oy_{E}^{\dot \a} { C}^{\b} 
+ \x^{\g \dot \d} \partial_{\g \dot \d}  \oF_E 
\\
\end{array}$}} 
\ee

\be
\la{brstransP} 
\vspace{.1in}
\framebox{{$\begin{array}{lll}  
 & &{\rm Nilpotent  \;Transformations \; for \; P}\\
\d P&= & 
\fr{\d {\cal A}}{\d \G_P} 
=  \y_{P  \b} {C}^{  \b} 
+ \x^{\g \dot \d} \partial_{\g \dot \d} P
\\
\d {\ov P} &= & 
\fr{\d {\cal A}}{\d {\ov \G}_P} 
=  {\ov \y}_{P  \dot \b} {\ov C}^{ \dot  \b} 
+ \x^{\g \dot \d} \partial_{\g \dot \d} {\ov P}
\\

\d \y_{P \a} &  =& \fr{\d {\cal A}}{\d {  Y}_P^{   \a} } = 
\pa_{ \a \dot \b } P {\ov C}^{\dot \b}  
+ 
C_{\a}   
F_P
+ \x^{\g \dot \d} \partial_{\g \dot \d}  \y_{P\a  }
\\

\d
 {\ov \y}_{P \dot \a} &  =& 
\fr{\d {\cal A}}{\d { {\ov Y}}_P^{\dot   \a} } = 
\pa_{ \a \dot \a }  {\ov P}_{} {C}^{\a}  
+ 
{\ov C}_{\dot \a}   
{\ov F}_{P}
+ \x^{\g \dot \d} \partial_{\g \dot \d} 
 {\ov \y}_{P \dot \a} 
\\
 \d F_P 
&=&\fr{\d {\cal A}}{\d \Lam_P} = 
  \pa_{\a \dot \b}   \y_P^{ \a} {\ov C}^{\dot \b} 
+ \x^{\g \dot \d} \partial_{\g \dot \d}  F_P 
\\
 \d \oF_P 
&=&\fr{\d {\cal A}}{\d \ov\Lam_P} = 
\pa_{\b \dot \a}   \oy_{P}^{\dot \a} { C}^{\b} 
+ \x^{\g \dot \d} \partial_{\g \dot \d}  \oF_P 
\\
\end{array}$}} 
\ee

\be
\la{brstransEPs} 
\vspace{.1in}
\framebox{{$\begin{array}{lll}  
& &{\rm Nilpotent  \;Transformations\; for\; E\; pseudofields}
\\
\d \G_E 
&= &
 \fr{\d {\cal A}}{\d E} 
=
  \Box    {\ov  E}_{} 
  +   m  F_P    
-\pa_{ \a \dot \b } Y_{E}^{ \a}    {\ov C}^{\dot \b}   
+ \x^{\g \dot \d} \partial_{\g \dot \d} \G_E
\\
\d {\ov \G}_E 
&= & \fr{\d {\cal A}}{\d {\ov E}} 
=
 \Box          { E} 
+  m   {\ov  F}_{ P}  -
 \pa_{ \a \dot \b } {\ov Y}_E^{ i \dot \b}    {C}^{\a}   
+ \x^{\g \dot \d} \partial_{\g \dot \d} 
 {\ov \G_E}
\\
\d Y_{E}^{ \a} 
&=&\fr{\d {\cal A}}{\d {  \y}_{ E  \a}} 
= 
-
  \pa^{\a \dot \b  }   
{\ov \y}_{  E \dot \b}
+ m
\y_P^{\a} 
-
\G_E  
 {C}^{  \a}
-
  \pa^{\a \dot \b}  \Lam_E   {\ov C}_{\dot \b} 
  + \x^{\g \dot \d} \partial_{\g \dot \d}  Y_{E}^{ \a}
\\
\d 
{\ov Y}^{ \dot \a} 
&=&\fr{\d {\cal A}}{\d {\ov \y}_E^{ \dot \a} 
} 
= 
-
  \pa^{\b \dot \a  }   
{ \y}_{ E\b}
+  m  
{\ov \y}_{P}^{\dot  \a} 
-
{\ov \G} _E 
 {\ov C}^{\dot  \a}
- 
  \pa^{\b \dot \a}  \ov\Lam_E   { C}_{\b} 
+ \x^{\g \dot \d} \partial_{\g \dot \d}  
{\ov Y}_E^{\dot \a} 
\\
 \d \Lam_E  
&=&  \fr{\d {\cal A}}{\d {F_E} } =
\oF_E + m P 
+ Y_E^{\a}
  C_{\a}+ 
\x^{\g \dot \d} \partial_{\g \dot \d}   \Lam_E  
\\
 \d {\ov \Lam}_E  
&=&  \fr{\d {\cal A}}{\d {\ov F_E} } =
F_E +m    \ov P 
+ \oY_E^{ \dot \a}
  \oC_{\dot\a}+ 
\x^{\g \dot \d} \partial_{\g \dot \d}   {\ov \Lam}_E   
\\
\end{array}$}} 
\ee

\be
\la{brstransPPs} 
\vspace{.1in}
\framebox{{$\begin{array}{lll}  
& &{\rm Nilpotent  \;Transformations\; for\; P\; pseudofields}
\\
\d \G_P 
&= &
 \fr{\d {\cal A}}{\d P} 
=
  \Box    {\ov  P}_{} 
  +   m  F_E    
-\pa_{ \a \dot \b } Y_{P}^{ \a}    {\ov C}^{\dot \b}   
+ \x^{\g \dot \d} \partial_{\g \dot \d} \G_P
\\
\d {\ov \G}_P 
&= & \fr{\d {\cal A}}{\d {\ov A}} 
=
 \Box          { P} 
+  m   {\ov  F}_{E}  -
 \pa_{ \a \dot \b } {\ov Y}_P^{  \dot \b}    {C}^{\a}   
+ \x^{\g \dot \d} \partial_{\g \dot \d} 
 {\ov \G_P}
\\
\d Y_{P}^{ \a} 
&=&\fr{\d {\cal A}}{\d {  \y}_{   \a}} 
= 
-
  \pa^{\a \dot \b  }   
{\ov \y}_{  P \dot \b}
+ m
\y_E^{\a} 
-
\G_P  
 {C}^{  \a}
- 
  \pa^{\a \dot \b}  \Lam_P   {\ov C}_{\dot \b} 
  + \x^{\g \dot \d} \partial_{\g \dot \d}  Y_{P}^{ \a}
\\
\d 
{\ov Y}^{\dot \a} 
&=&\fr{\d {\cal A}}{\d {\ov \y}_P^{ \dot \a} 
} 
= 
-
  \pa^{\b \dot \a  }   
{ \y}_{ P\b}
+  m  
{\ov \y}_{E}^{\dot  \a} 
-
{\ov \G} _P 
 {\ov C}^{\dot  \a}
- 
  \pa^{\b \dot \a}  \ov\Lam_P   { C}_{\b} 
+ \x^{\g \dot \d} \partial_{\g \dot \d}  
{\ov Y}_P^{\dot \a} 
\\
 \d \Lam_P  
&=&  \fr{\d {\cal A}}{\d {F_P} } =
\oF_P + m E 
+ Y_P^{\a}
  C_{\a}+ 
\x^{\g \dot \d} \partial_{\g \dot \d}   \Lam_P  
\\
 \d {\ov \Lam}_P  
&=&  \fr{\d {\cal A}}{\d {\ov F_P} } =
F_P +m    \ov E 
+ \oY_P^{ \dot \a}
  \oC_{\dot\a}+ 
\x^{\g \dot \d} \partial_{\g \dot \d}   {\ov \Lam}_P   
\\
\end{array}$}} 
\ee

\be
\vspace{.1in}
\framebox{{$\begin{array}{lll}  
 & &{\rm Nilpotent \; CDSS\;Transformations}\\
\d \f_{\dot\a}&= & 
\fr{\d {\cal A}}{\d G^{\dot\a}} 
=  (C^{\d} X_{ \d\dot \a} )
 \\
\d \f^{\dot\a}&= & 
\fr{\d {\cal A}}{\d G_{\dot\a}} 
=  (- C_{\d} X^{ \d\dot \a})
 \\
\d X_{ \a\dot \b}  &  =& 
\fr{\d {\cal A}}{\d  \S^{ \a \dot \b} } = 
(\pa_{ \a \dot \d }  \f_{\dot\b} {\ov C}^{\dot \d}  
+ 
C_{\a}   
\c_{\dot\b})
 \\
\d X^{\b\dot \a }  &  =& 
\fr{\d {\cal A}}{\d  \S_{\b \dot \a } } = 
(-\pa^{ \b \dot \d }  \f^{\dot\a} {\ov C}_{\dot \d}  
+ 
C^{\b}   
\c^{\dot\a})
 \\
 \d \c_{\dot\a}
&=&\fr{\d {\cal A}}{\d L^{\dot \a}} = 
(  \pa^{\d\dot \d}  X_{ \d \dot \a}   {\ov C}_{\dot \d} )
 \\
 \d \c^{\dot\a} 
&=&\fr{\d {\cal A}}{\d L_{\dot \a}} = 
(  \pa_{\d \dot \d}  X^{ \d\dot \a }   {\ov C}^{\dot \d} )
 \\
\end{array}$}} 
\ee

\be
\vspace{.1in}
\framebox{{$\begin{array}{lll}  
 & &{\rm Nilpotent \; CDSS\;Pseudofield\;Transformations}\\
\d G^{\dot\a}&= & 
\fr{\d {\cal A}}{\d \f_{\dot\a}} 
=  {\rm etc. \;are\; not\; needed\;here\;for\;now}
\\
\end{array}$}} 
\ee

   \section{Completion of the action that includes the \ei\ $\cA_{X}$: the completed action $\cA_{\rm Completed} $ }
   \la{completionsection}
  Since we know that the equation (\ref{constraintoncAX}) 
  is true, we want to add the \ei s $g \cA_{X}+\og {\ov \cA}_{X}$ to the above action $\cA$   with new coupling contants $g,\og$,   
The natural question is whether this new action  $ \cA_1$  satisfies the same \ME, described in section (\ref{mastereqsection}), that the  action $ \cA$ did.  Clearly there is no reason to expect this.  However we conjecture that
the action $  \cA_{\rm Completed}$, of the form 
    \be
  \cA_{\rm Completed} =  \cA + g \cA_{X}+\og {\ov \cA}_{X}+g^2 \cA_{\rm X,2}+\og^2 {\ov \cA}_{\rm X,2}
  +g \og {\cA}_{\rm X,2,Mixed}
 \la{treeaction} \ee
does in fact satisfy the same \ME, as defined in section \ref{mastereqsection}.  In other words we expect that $\cA_{\rm Completed} $ can be substituted for $\cA$ in equation (\ref{mastereq}) and its components (\ref{mastereq1}), (\ref{mastereq2}), (\ref{mastereq3}) and (\ref{masterstructure}).

In the formula (\ref{treeaction}) we define
\be
 \cA_{\rm X,2}=
   \cA_{\rm X, 2,E} -  \cA_{\rm X, 2,P}
 \ee
   where
   \be
 \cA_{\rm X, 2,E} =   \int d^4 x     \lt \{
b_{12} \f^{\dot \a}\f_{\dot \a} 
 F_E \oE  +b_{13} \f^{\dot \a} X_{ \b\dot \a}
 \y_E^{\b}  \oE   \ebp
 +  b_{14}X^{\b\dot \a } X_{\b\dot \a } E   \oE  + b_{15} \f^{\dot \a} \c_{\dot \a} E \oE  
\rt \} 
  \ee
 and  
 \be
 \cA_{\rm X,2,Mixed}=
   \cA_{\rm X, 2,Mixed,E} -  \cA_{\rm X, 2,Mixed,P}
 \ee  
where\be
  \cA_{\rm X, 2,Mixed,E}   =   \ov \cA_{\rm X, 2,Mixed,E}   =   \int d^4 x     \lt \{
b_{16} \f^{\dot \a}\ov \f^{\b}   E \pa_{\b \dot \a} \oE
    +b_{17} \f^{\dot \a}\ov \f^{\b}   \y_{E \b} \oy_{E \dot\a}
  \ebp  +b_{18} \f^{\dot \a} \ov X_{\b \dot \a }
 \y_E^{\b}  \oE   +b_{19} \ov \f^{ \b}   X_{\b\dot \a }
E \oy_E^{\dot \a}  
 +  b_{20}X^{\b \dot \a } \ov X_{ \b \dot \a} E   \oE  
\rt \} 
  \ee

We need to ensure that the action $\cA_{\rm Completed} $  is real.  This looks reasonable, given the forms of the above terms.  Judging from the work that was needed to find the coefficients 
(\ref{valsofb}), there will be quite a lot of very similar  work to find the coefficients 
$ b_{12}, \cdots b_{20} $ that are needed to complete the action $\cA_{\rm Completed} $ 
in this section \ref{completionsection}. The necessary detail to accomplish that, and to find the values of $b_i$ in (\ref{valsofb}) will be discussed in future papers. We will refer to the terms $g^2 \cA_{\rm X,2}+\og^2 {\ov \cA}_{\rm X,2}
  +g \og {\cA}_{\rm X,2,Mixed}$ in (\ref{treeaction}) as the Completion Terms.  The existence of these terms is not guaranteed by the spectral sequence however, and this conjecture, which will be shown to be correct in \ci{E10}, says something new (and not well understood at present) about the cohomology here.

\section{Conclusion}

In this paper E4, we have provided the basic structure necessary to verify that the expression (\ref{theEI}) is invariant under the transformations in section \ref{brsoperatorsection}.   The main new feature here, which was not present in the simpler example in E3, is that here we need to cancel the variations (\ref{variationsofXEandXP1}) and (\ref{variationsofXEandXP2})  between the two very similar exotic expressions in (\ref{theEI}).  This is the essence of the solution of the constraint that arises in the spectral sequence in \ci{E2}, and it is crucial for the XM in \ci{E6}.  The other new feature here in E4  is the conjecture for the form of $\cA_{\rm Completed} $  that is contained in section \ref{completionsection}. Both of these features require a computer program to verify the claims here, and to make the calculation of the coefficients a reasonable task, and that will be presented in \ci{E10}. In \ci{E5} we will return to this EP model and add a U(1) gauge interaction.  It turns out that this is not very difficult--it just changes a few terms.  But those terms are very important, as we will see in E6.

\begin{center}
 { Acknowledgments}
\end{center}
\vspace{.1cm}

  I thank    Howard Baer, Friedemann Brandt, Philip Candelas,   Mike Duff, Sergio Ferrara,  Dylan Harries, Marc Henneaux,  D.R.T. Jones, Olivier Piguet, Antoine van Proeyen,     Peter West and Ed Witten for stimulating correspondence and conversations.   I also express appreciation for help in the past from William Deans, Lochlainn O'Raifeartaigh, Graham Ross, Raymond Stora, Steven Weinberg, Julius Wess and Bruno Zumino. They are not replaceable and they are missed.  I also  thank Ben 
Allanach, Doug Baxter,  Margaret Blair,  Murray Campbell, David Cornwell, Thom Curtright, James Dodd, Richard Golding, Chris T.  Hill,  Davide Rovere,   Pierre Ramond, Peter Scharbach,  Mahdi Shamsei, Sean Stotyn, Xerxes Tata and J.C. Taylor, for recent, and helpful, encouragement to carry on with this work. I also express appreciation to Dylan Harries and  to Will, Dave and Peter Dixon, Vanessa McAdam and Sarunas Verner for encouraging and teaching me to use coding.  I note with sadness the recent passing of Carlo Becchi and Kelly Stelle, both of whom were  valuable colleagues and good friends.


 \tiny 
\articlenumber\\
Feb 4, 2026
\end{document}